\documentclass{article}
\usepackage{arxiv}

\usepackage{amsmath}

\usepackage{amssymb}
\usepackage{latexsym}

\usepackage{url}
\usepackage{xcolor}

\usepackage{hyperref}
\usepackage{mathtools}

\usepackage{multirow}

\usepackage{algorithm}
\usepackage[noend]{algpseudocode}

\title{Privacy-Preserving Product-Quantized Approximate Nearest Neighbor Search Framework for Large-scale Datasets via A Hybrid of Fully Homomorphic Encryption and Trusted Execution Environment}

\author{
	Shozo Saeki \\
	Center for Information Technology \\
	Ehime University \\
	Matsuyama, Ehime and 790-8577, Japan \\
	\texttt{saeki.shozo.cg@ehime-u.ac.jp} \\
	\And
	Minoru Kawahara \\
	Center for Information Technology \\
	Ehime University \\
	Matsuyama, Ehime and 790-8577, Japan \\
	\texttt{kawahara@ehime-u.ac.jp} \\
	\And
	Hirohisa Aman \\
	Center for Information Technology \\
	Ehime University \\
	Matsuyama, Ehime and 790-8577, Japan \\
	\texttt{aman@ehime-u.ac.jp} \\
}

\renewcommand{\shorttitle}{PPPQ-ANN}

\begin{document}
\maketitle

\begin{abstract}
A nearest-neighbor framework is a fundamental tool for various applications involving Large Language Models (LLMs) and Visual Language Models (VLMs).
Vectors used for nearest-neighbor searches have richer information for similarity searches.
This information leads to security risks, such as embedding inversion and membership attacks.
Therefore, Privacy-Preserving Approximate Nearest-Neighbor (PP-ANN) approaches are necessary for highly confidential data.
However, conventional PP-ANN approaches based on a Trusted Execution Environment (TEE) or Fully Homomorphic Encryption (FHE) do not achieve practical security or performance.
Additionally, conventional approaches focus on the search process rather than database generation for nearest-neighbor.
To address these issues, we propose a Privacy-Preserving Product-Quantization Approximate Nearest Neighbor (PPPQ-ANN) framework.
PPPQ-ANN provides a multi-layered security structure for vectors based on a hybrid of FHE and TEE.
Additionally, PPPQ-ANN minimizes FHE ciphertext computations by combining Product-Quantization (PQ) with optimized data packing.
We demonstrate the performance of PPPQ-ANN on million-scale datasets.
As a result, PPPQ-ANN achieves database generation in less than 2 hours and more than 50 QPS in a sequential search while preserving privacy.
Therefore, PPPQ-ANN optimizes the trade-off between security and performance by utilizing a hybrid of FHE and TEE, achieving practical performance while preserving privacy.
\end{abstract}



\section{Introduction}
With advances in machine learning, such as large language models (LLMs) \cite{RAG, BlendedRAG} and visual language models (VLMs) \cite{CLIP, GeminiEmbedding}, these models can achieve high-quality retrievals.
While various retrieval methods exist, nearest-neighbor search plays a fundamental role among them \cite{RAG, BlendedRAG, CLIP, GeminiEmbedding}.
The vectors used in nearest-neighbor search contain rich information that enables similarity-based retrieval.
These vectors pose risks to data confidentiality and privacy \cite{SentenceInversion, ZeroShotTextInversion, DNAInversion, MembershipAttack, AttributionAttack}.
These risks include attacks such as embedding inversion, membership inference, and attribute inference \cite{InfoLeak, TransferInversion, MembershipAttack, AttributionInferenceAttack}.
Therefore, these vectors, especially high-confidentiality data, require a way to preserve data confidentiality and privacy against attacks.

Representative approaches to confidential computation are Multi-Party Computation (MPC) \cite{MPC_Survey, MPCBOOK}, Trusted Execution Environment (TEE) \cite{TEE}, and Fully Homomorphic Encryption (FHE) \cite{TFHE, CKKS}.
Various methods have been proposed for Privacy-Preserving Approximate Nearest Neighbor (PP-ANN) search that aim to protect privacy by leveraging these confidential computations \cite{SANNS, TEEANN, PPAkNN, GraSS}.
MPC-based and TEE-based PP-ANN methods achieve practical performance.
However, they have inherent security limitations, including the leakage of access patterns or the risk of side-channel attacks \cite{TEESurvey, MPC_Survey}.
On the other hand, FHE-based methods achieve strong cryptographic guarantees without relying on hardware trust.
However, they face the extremely high computational cost of large-scale ANN search \cite{SANNS, PPAkNN}.
These issues make them impractical in latency-sensitive applications.
Additionally, conventional PP-ANN approaches focus on the security of the search process.
However, the database generation processes for PP-ANN are not proposed, and they are also important for privacy-preserving applications.

\begin{figure}[tbp]
    \begin{center}
        \includegraphics[width=0.5\hsize]{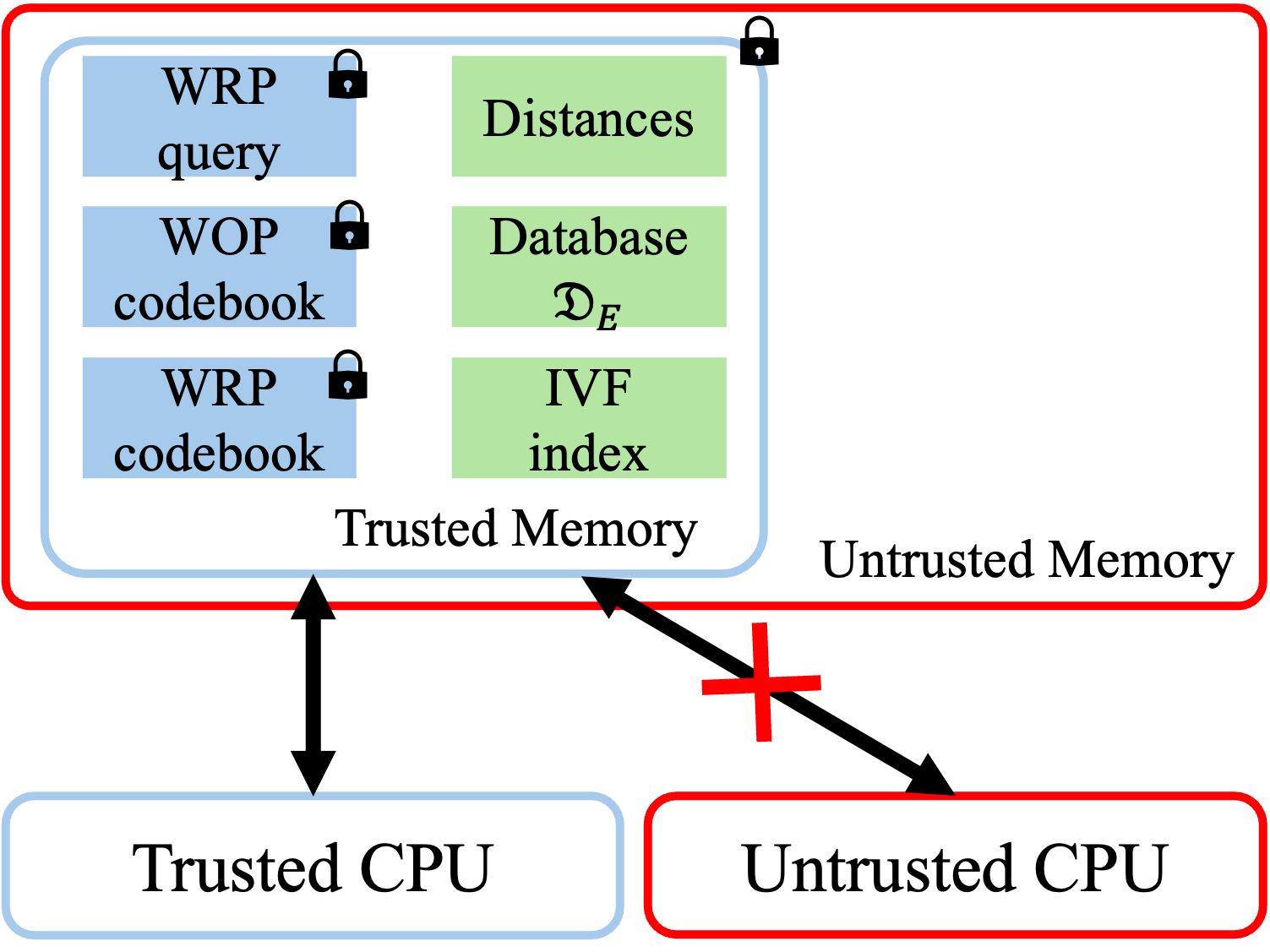}
    \end{center}
    \caption{
        Server memory of the search process in PPPQ-ANN.
        All processing in PPPQ-ANN is performed within the TEE, and data is memorized in encrypted RAM.
        Blue squares denote ciphertexts encrypted by CKKS-based FHE, and green squares denote plaintexts.
        Queries and codebooks that could lead to the complete reconstruction of data are encrypted within the TEE using CKKS-based FHE.
        WOP and WRP denote optimized data packings for PQ.
        Additionally, database $\mathcal{D}_{E}$ and IVF index are represented by PQ code indexes.
    }
    \label{fig:PPPQANNFHETEE}
\end{figure}

To address these issues, this paper proposes a Privacy-Preserving Product-Quantization Approximate Nearest Neighbor (PPPQ-ANN) framework. 
PPPQ-ANN is a hybrid approach that combines CKKS-based FHE and TEE, and employs multi-layered security.
PPPQ-ANN provides a privacy-preserving method for both database generation and search processes.
Additionally, all processes of PPPQ-ANN are performed on TEE.
However, TEE cannot provide complete privacy-preserving processes.
Then, PPPQ-ANN encrypts the original database, codebooks, and queries using CKKS-based FHE to preserve privacy.
PPPQ-ANN uses CKKS data packings optimized for Product Quantization (PQ).
Thus, PPPQ-ANN provides multi-layered security for more sensitive data, such as medical and financial data.
Figure~\ref{fig:PPPQANNFHETEE} shows multi-layered security for data.
Finally, we demonstrate the PPPQ-ANN performances on several million-scale datasets.
The findings of this paper are as follows:
\begin{itemize}
    \item We propose PPPQ-ANN, which is a hybrid framework of CKKS-based FHE and TEE for all processes of ANN. PPPQ-ANN provides multi-layered security beyond standalone FHE or TEE. Unlike conventional FHE-only approaches, a hybrid framework optimizes the trade-off between cryptographic strength and computational efficiency by leveraging TEE for performance-critical operations while maintaining privacy-preserving computation through FHE.
    \item By PQ with optimized data packings for PQ, PPPQ-ANN significantly reduces the high computational cost and network traffic of FHE in both database generation and search processes, while maintaining high search accuracy (Recall@10 $>$ 0.9).
    \item PPPQ-ANN achieves less than 2 hours for the database generation process and more than 50 QPS for the sequential search process. These results demonstrate that PPPQ-ANN can be executed in a practical time.
\end{itemize}

\section{Related Work}
Various approaches have been proposed for secure machine learning.
These include multi-party computation (MPC) \cite{CrypTen, MPC_CNN, MPC_Survey}, trusted execution environments (TEEs) \cite{GuaranTEE, FLTEE, PliniusTEE}, and fully homomorphic encryption (FHE) \cite{SecurekMeans,SecureKMCMC,FHENN}.
All of these aim to enhance security and preserve data privacy.
MPC enables multiple parties to jointly compute various tasks without sharing original data with one another \cite{MPCBOOK}.
TEE performs computations in a trusted environment, where it is isolated and independent of other systems \cite{TEE}.
FHE performs computations on encrypted data, and servers cannot access the plaintexts \cite{CKKS}.
Practical machine learning applications have been developed using MPC and TEE \cite{MPC_CNN,PliniusTEE}.
However, MPC and TEE have known security limitations.
In MPC, the presence of malicious participants in the computation causes a risk of decreased computational efficiency and security vulnerabilities \cite{MPCBOOK}.
TEE is vulnerable to side-channel attacks, and there is a risk of data leakage in a trusted environment \cite{TEEAtack}.
On the other hand, FHE is one of the Post-Quantum Cryptographic (PQC) schemes and has higher security for preserving data confidentiality than TEE \cite{TEEvsFHE}.
However, FHE has a significantly higher computational cost than MPC and TEE.
Some approaches have been proposed as secure machine learning over FHE \cite{TPClustering, SecurekMeans, FHENN, SecureKMCMC}.
Most of these \cite{TPClustering,SecurekMeans, FHENN} suffer from low scalability due to high computational costs.
Thus, machine learning over FHE is necessary for reducing computational cost.

FHE was initially proposed by Gentry in 2009 \cite{FirstFHE} using ideal lattice constructions.
Gentry's FHE suffers from drawbacks, including high computational cost and noise control.
Next, Brakerski-Gentry-Vaikuntanathan (BGV) \cite{BGV}, Brakerski-Fan-Vercauteren (BFV) \cite{BFV}, and Gentry-Sahai-Waters (GSW) \cite{GSW} were proposed using Learning With Errors (LWE) and Ring-LWE (RLWE).
These FHE schemes support operations over integer-valued data.
Subsequently, Cheon-Kim-Kim-Song (CKKS) \cite{CKKS} was proposed, which can operate with real-valued data.
The computation of CKKS can be performed using Single Instruction Multiple Data (SIMD) operations.
However, CKKS cannot directly compute the non-linear functions.
Thereafter, Fully Homomorphic Encryption over the Torus (TFHE) was proposed \cite{TFHE}.
TFHE is optimized for handling bit-level data.
Most modern FHE schemes support data packing, which packs multiple data values into a ciphertext \cite{BGV, CKKS, TFHE}.
This feature is critical for enhancing computational efficiency and reducing network traffic \cite{SecurekMeans,SecureKMCMC}.

Some approaches have been proposed for privacy-preserving approximate nearest neighbor (PP-ANN) \cite{SANNS, TEEANN, PPAkNN, GraSS}.
PP-ANN with TEEs \cite{TEEANN} is fast because it executes with plaintext within TEEs.
However, TEEs rely heavily on hardware reliability and have a weakness in security \cite{TEESurvey}.
Therefore, PP-ANN with TEEs is weak at handling highly confidential data.
On the other hand, PP-ANN with FHE \cite{SANNS, PPAkNN} achieves higher levels of confidentiality than one with TEEs.
However, these suffer from high computational cost and are less than 1 QPS (queries per second) \cite{SANNS, PPAkNN}.

\section{Privacy-Preserving Product-Quantization Approximate Nearest Neighbor Framework}
\begin{figure}[tbp]
    \begin{center}
        \includegraphics[width=0.9\hsize]{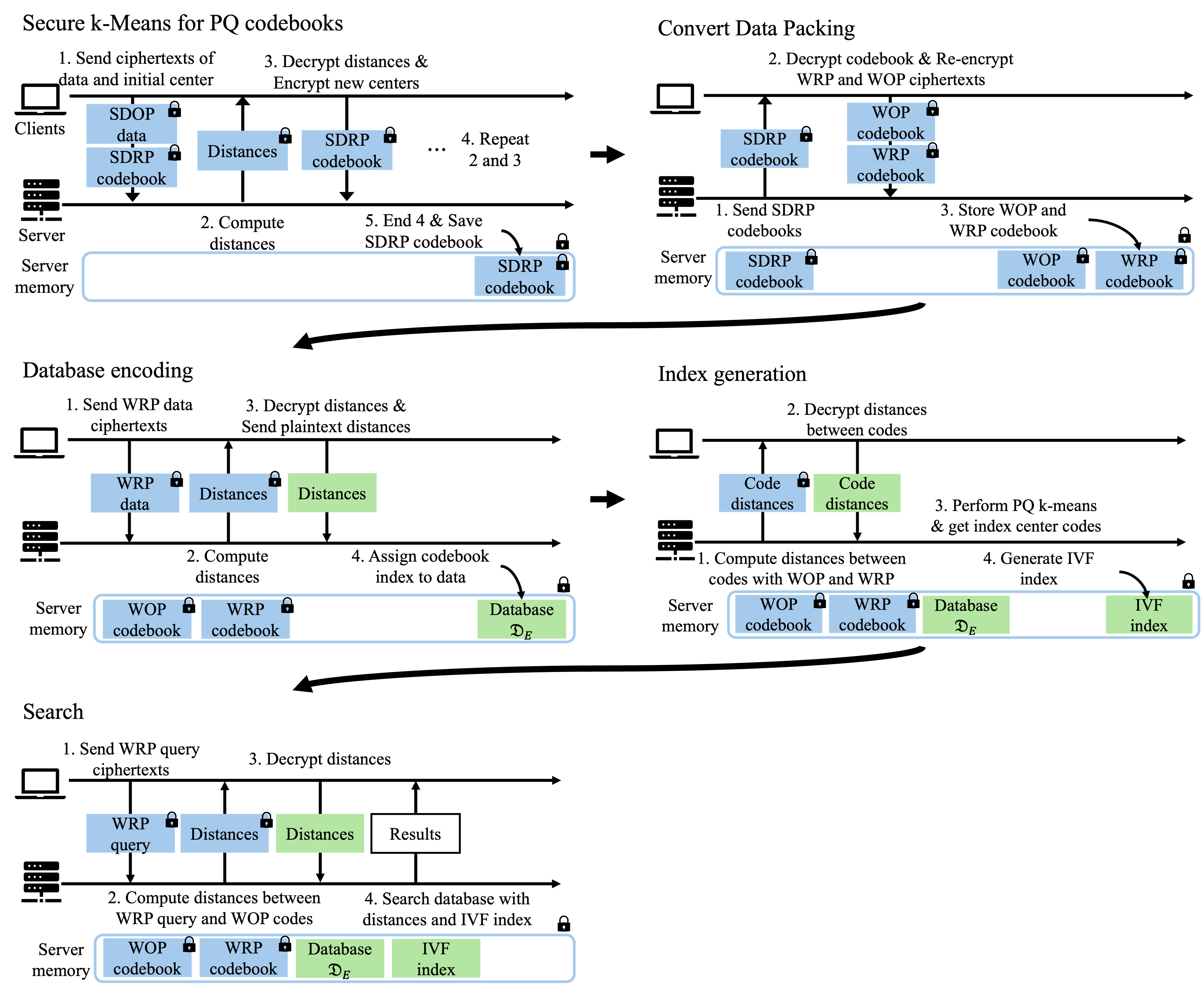}
    \end{center}
    \caption{
        PPPQ-ANN processing flow.
        PPPQ-ANN can be broken down into four parts: codebook generation (convert data packing), database encoding, index generation, and search.
        Each part shows the communication between clients and server, and the summary of proccesses.
        Blue squares denote ciphertexts encrypted by CKKS-based FHE, and green squares denote plaintexts.
        SDOP, SDRP, WOR, and WRP denote kinds of data packing.
        In all processes, server memory is encrypted by TEE.
    }
    \label{fig:SPQANNflow}
\end{figure}

This section proposes a Privacy-Preserving Product-Quantization Approximate Nearest Neighbor (PPPQ-ANN) framework.
PPPQ-ANN combines product quantization (PQ) with the CKKS scheme, enabling practical use of large-scale datasets while maintaining low computational complexity and minimal network traffic based on FHE ciphertexts.
PPPQ-ANN comprises of four parts: codebook generation for product quantization (PQ), database encoding, database indexing, and search.
Figure~\ref{fig:SPQANNflow} shows the PPPQ-ANN flow.
Additionally, all PPPQ-ANN server processes are performed in TEE.
Even within the TEE's memory, data, codebooks, and queries that could be used to restore the original data are encrypted using CKKS.
On the other hand, distances and indices that cannot strictly restore the original data are preserved only by TEE.
Thus, PPPQ-ANN provides a multi-layered security for all ANN processes, as shown in Figure~\ref{fig:PPPQANNFHETEE}.
This structure optimizes the trade-off between security and computational cost.
The threat model of PPPQ-ANN is an honest-but-curious.

First, we propose the data packing for the CKKS scheme in PPPQ-ANN.
Then, we propose an efficient algorithm for computing distances during data packing.
Next, we propose codebook generation, database encoding, database indexing, and a search algorithm.
Finally, we consider the computational complexity and network traffic.
For simplicity, this paper assumes a single client and server when explaining the algorithms.

\subsection{Data Packing for Product Quantization}
Ciphertexts encrypted by the CKKS scheme can pack $n_{p} = \frac{N_{r}}{2}$ floating points, where $N_{r}$ denotes the ring dimension of the CKKS scheme.
Let $\mathcal{D} = \left\{ \mathbf{x}_{i} \right\}_{i=1}^{N_{\mathcal{D}}} \in \mathbb{R}^{N_{\mathcal{D}} \times d}$ denote the data, where $N_{\mathcal{D}}$ denotes the number of data points and $d$ denotes the data dimension.
PPPQ-ANN actually packs data divided by subspace into ciphertext because PPPQ-ANN uses the PQ \cite{PQ}.
Let $\mathcal{D}_{SD} = \left\{ \mathbf{x}_{i_d, i_s} \right\}_{i_d=1, i_s=1}^{N_{\mathcal{D}}, n_{s}} \in \mathbb{R}^{N_{\mathcal{D}} \times n_{s} \times d_{s}}$ denote the divided data, where $n_{s}$ denotes the number of the subspacs and $d_{s} = \lceil\frac{d}{n_{s}}\rceil$ denotes the dimension of subspaces data.
Additionally, let $\mathcal{C} = \left\{ \mathbf{c}_{i_c, i_s} \right\}_{i_{c}=1, i_s=1}^{N_{C}, n_{s}} \in \mathbb{R}^{N_{C} \times n_{s} \times d_{s}}$ denote the PQ's codebooks, where $N_{C}$ denotes the codebook size for each subspace.
Note that if $d$ is not divisible by $n_{s}$, PPPQ-ANN performs a zero fill.
For example, if data $\{ x_1, x_2, x_3, x_4 \}$ is divided by 3 subspaces, sub data are $\{ x_1, 0 \}$, $\{ x_2, 0 \}$, and $\{ x_3, x_4 \}$.
This division is equivalent to considering dimension $d$ as pseudo-dimension $d^{\prime} = n_{s} \lceil \frac{d}{n_{s}} \rceil$. 

\begin{figure}[tbp]
    \begin{center}
        \includegraphics[width=\hsize]{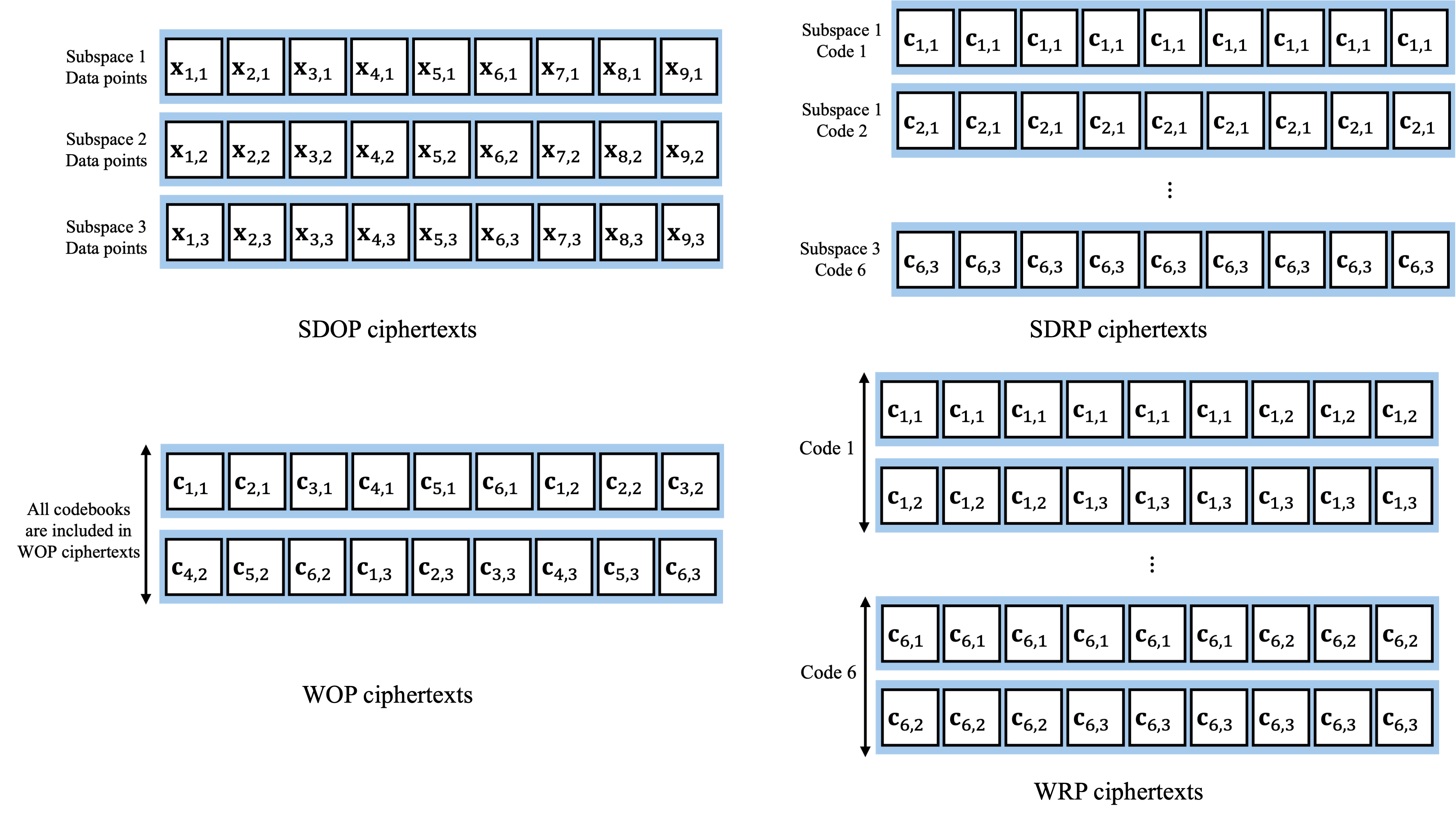}
    \end{center}
    \caption{
        Four types of data packing examples.
        Blue squares represent a ciphertext.
        This example has 3 subspaces, 9 data points, and 6 codes for each subspace.
        SDOP ciphertexts pack 9 data points for each subspace.
        SDRP, WOP, and WRP pack 18 (3 subspace $\times$ 6 codes) codes as a codebook.
    }
    \label{fig:DataPackingTypes}
\end{figure}

PPPQ-ANN uses four types of data packing based on order packing (OP) and repeated packing (RP) for the CKKS scheme \cite{SecurekMeans, SecureKMCMC}.
OP packs data into an array in order.
RP repeatedly packs one piece of data into the array up to the same length as the corresponding OP.
Four types of data packing are subdimension-wise order packing (SDOP), subdimension-wise repeated packing (SDRP), whole-order packing (WOP), and whole-repeated packing (WRP).
Data are packed into ciphertexts using SDOP or WRP.
On the other hand, PQ codebooks are packed into ciphertexts using SDRP, WOP, or WRP.
Figure~\ref{fig:DataPackingTypes} shows the four types of data packing examples.
The difference between subdimension-wise packings (SDOP and SDRP) and whole packings (WOP and WRP) is whether the ciphertext is divided by subspace.
SDOP and SDRP are simply OP and RP in every subdimension.
On the other hand, when WOP finishes packing one subspace of data, it packs the next subspace of data.
WRP packs subspace data corresponding to WOP subspace data.
Hence, WOP and WRP can reduce the number of ciphertexts compared to SDOP and SDRP.
SDOP and SDRP are only used in codebook generation process.
On the other hand, WOP and WRP are used in database encoding, indexing, and search processes.
The maximum number of subdimension data per ciphertext $n_{pd}$ is the same for all four types of data packing and follows as:
\begin{equation}
n_{pd} = \left\lceil \frac{n_{p}}{d_{s}} \right\rceil = \left\lceil \frac{N_{r}}{2 \left\lceil \frac{d}{n_{s}} \right\rceil} \right\rceil.
\end{equation}

\subsection{Faster Distance Computation}
Distance computations in PPPQ-ANN perform asymmetric distance computation (ADC) \cite{PQ} on ciphertexts.
Note that the codebook generation process simply utilizes the distances computed between ciphertexts in each subdimension.

We propose an algorithm that performs distance computation for each subdimension $d_{s}$ in SDOP, SDRP, WOP, and WRP.
This distance computation algorithm is performed between OP-based and RP-based ciphertexts.
Note that distance computation between SDOP and SDRP and between WOP and WRP can be performed in this paper.
Ciphertext computations of PPPQ-ANN are only distance computations.
The other computations can be performed in plaintext.
Ciphertext computations take much longer than plaintext computations.
Hence, ciphertext distance computations are crucial for the performance of PPPQ-ANN.
A conventional distance computation method was proposed using slot-wise arithmetic operations and a ciphertext summation algorithm \cite{SecurekMeans}.
This summation algorithm uses a ciphertext rotation function ($\mathrm{rot} \left( \cdot, \cdot \right)$).
The number of rotations is crucial for the performance of distance computations because it increases with the size of the computation dimension $d_{s}$.
The other operations, slot-wise arithmetic operations, are not 
The complexity of conventional summation algorithms \cite{SecurekMeans} is $O ( \lfloor \log_{2} d_{s} \rfloor + d_{s} - 2^{\lfloor \log_{2} d_{s} \rfloor})$.
Conventional summation algorithms are only fast when the computation dimension $d_{s}$ is a power of 2.

\begin{figure}[tbp]
    \begin{center}
        \includegraphics[width=0.5\hsize]{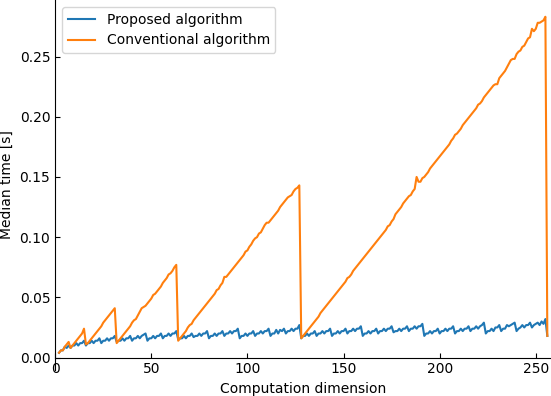}
    \end{center}
    \caption{
        The relationship between data dimension and computation time for Euclidean distance computations over the CKKS scheme on the Apple M2 Max chip.
        The computation time for each dimension is the median time among 100 computations.
        The blue line shows the results of the proposed algorithm.
        The orange line shows the results of the conventional algorithm \cite{SecurekMeans}.
        The computational time of the proposed method increases only logarithmically.
    }
    \label{fig:CompTimeDimension}
\end{figure}

We propose a fast summation algorithm for all computation dimensions $d_{s}$.
The summation algorithm is shown in Algorithm~\ref{alg:RotSummation}.
\begin{algorithm}[tb]
    \caption{Faster summation algorithm for CKKS ciphertexts}
    \label{alg:RotSummation}
    \textbf{Input:} Ciphertext $c$ and the number of dimension $d_{s}$ \\
    \textbf{Output:} Ciphertext $r$ containing the summation results
    \begin{algorithmic}[1]
        \State {$i \leftarrow 0$}
        \State {$j \leftarrow 0$}
        \State {$r \leftarrow c$}
        \State {$r_{m} \leftarrow$ initialize the ciphertexts array, length is $\lfloor \log_{2} d_{s} \rfloor$}
        \If {$d_{s} \neq 1$}
            \While {$i < \lfloor \log_{2} \left( d_{s} \right)  \rfloor$}
                \State {$r \leftarrow r + \mathrm{rot} \left( r, 2^{i} \right)$}
                \State {$r_{m} [ i ] \leftarrow r$}
                \State {$i \leftarrow i + 1$}
            \EndWhile
            \While {$d_{s} - 2^{i} - j \geq 2$}
                \State {$k \leftarrow \lfloor \log_{2} (d_{s} - 2^{i} - j) \rfloor$}
                \State {$r \leftarrow r + \mathrm{rot} \left( r_{m}[k-1], 2^{i}+j \right)$}
                \State {$j \leftarrow j + 2^{k}$}
            \EndWhile
            \If {$d_{s} \mod 2 = 1$}
                \State {$r \leftarrow r + \mathrm{rot} \left( c, 2^{i}+j \right)$}
            \EndIf
        \EndIf
    \end{algorithmic}
\end{algorithm}
This algorithm memorizes the results of the first loop rotation summation.
Then, the second loop utilizes these memorized results.
These loops require $n_{rot} < 2 \log_{2} d_{s}$ iterations of the rotation function and addition for ciphertexts.
This algorithm can compute the summation with computational complexity as follow:
\begin{equation}
    \label{eq:DistanceComplexity}
    O \left( n_{rot} \right) = O \left( 2 \log_{2} d_{s} \right).
\end{equation}
In addition, spatial complexity is $O \left( \log_{2} d_{s} \right)$.
This algorithm needs a bit more spatial complexity than a conventional algorithm.
However, this algorithm can significantly reduce the number of rotation functions.
For instance, when $d_{s} = 100$, the number of rotation functions is 42 and 8 for the conventional algorithm \cite{SecurekMeans} and Algorithm~\ref{alg:RotSummation}, respectively.
Finally, Figure~\ref{fig:CompTimeDimension} shows the relationship between data dimension and computation time for Euclidean distance computations on the server.

\subsection{Codebook Generation}
Codebook generation utilizes the k-means algorithm with the CKKS scheme encryption.
Additionally, codebook generation involves converting the SDRP codebook to WOP and WRP codebooks after generating a codebook for each subspace.
One approach has been proposed for secure k-means \cite{SecurekMeans}.
This approach performs comparison operations at a server algorithm using polynomial approximation.
However, this approximation requires more multiplicative depth, which in turn makes ciphertexts larger.
Large ciphertexts could lead to a slowdown and much network traffic.
Therefore, we propose a simple and secure k-means algorithm for codebook generation.
Note that PPPQ-ANN uses random sampling or the secure k-MCMC \cite{SecureKMCMC} as initialization algorithm.

Secure k-means uses SDOP and SDRP for data and code packing, respectively.
Data $\mathcal{D}_{RS} \in \mathbb{R}^{N_{RS} \times d_{s}}$ are randomly sampled from each subspace of $\mathcal{D}_{SD}$, where $N_{RS}$ means the number of randomly sampled data.
Data ciphertexts $C_{SDOP}$ pack $N_{RS}$ data using SDOP, and code ciphertexts $C_{SDRP}$ pack $N_{C}$ codes using SDRP.
Therefore, the number of data ciphertexts $N_{SDOP}$ and code ciphertexts $N_{SDRP}$ for each secure k-means are as follows:
\begin{align}
N_{SDOP} &= \left\lceil \frac{N_{RS}}{n_{pd}} \right\rceil, \\ 
N_{SDRP} &= N_{C}.
\end{align}
The server and client algorithms for secure k-means are presented in Algorithms~\ref{alg:ServerKMeans} and \ref{alg:ClientKMeans}, respectively.
In these algorithms, a server can only know the labels.
\begin{algorithm}[tb]
    \caption{Server algorithm of secure k-means}
    \label{alg:ServerKMeans}
    \textbf{Input:} Data ciphertexts $C_{SDOP}$, initial code ciphertexts $C_{SDRP}$, the number of data $N_{RS}$, and the maximum iteration of k-means $n_{k}$ \\
    \textbf{Output:} Code ciphertexts $C_{WOP}$ and $C_{WRP}$ encrypted with WOP and WRP
    \begin{algorithmic}[1]
        \State {$l_{p} \leftarrow$ initialize the previous label array to $\mathbf{0}$, length is $N_{RS}$}
        \State {$l_{c} \leftarrow$ initialize the current label array to $\mathbf{1}$, length is $N_{RS}$}
        \State {$i \leftarrow 1$}
        \While {$i \leq n_{k}$}
            \If {Check if $l_{c}$ is the same as $l_{p}$}
                \State {Break the while loop}
            \Else
                \State {Copy $l_c$ to $l_p$}
            \EndIf
            \State {$C_{D} \leftarrow$ compute the distances between $C_{SDOP}$ and $C_{SDRP}$ with Algorithm~\ref{alg:RotSummation}}
            \State {Send distance ciphertexts $C_{D}$ to client}
            \State {Recieve new code ciphertexts $C_{SDRP}$ and labels $l_{c}$ from client with Algorithm~\ref{alg:ClientKMeans}}
            \State {$i \leftarrow i + 1$}
        \EndWhile
        \State {Send $C_{SDRP}$ to client and receive $C_{WOP}$ and $C_{WRP}$ converted from $C_{SDRP}$}
    \end{algorithmic}
\end{algorithm}
\begin{algorithm}[tb]
    \caption{Client algorithm of secure k-means}
    \label{alg:ClientKMeans}
    \textbf{Input:} Original data $D_{R}$
    \begin{algorithmic}[1]
        \State {Receive distance ciphertexts $C_{D}$ from server}
        \State {$D \leftarrow$ decrypt $C_{D}$}
        \State {$l_{c} \leftarrow$ assign the label to $D_{R}$ from $D$}
        \State {$C \leftarrow$ compute new $N_{C}$ codes based on $l_{c}$ and $D_{R}$}
        \State {$C_{SDRP} \leftarrow$ encrypt new codes with SDRP}
        \State {Send $C_{SDRP}$ and $l_{c}$ to server}
    \end{algorithmic}
\end{algorithm}

Finally, codebook generation converts SDRP codebooks to WOP and WRP codebooks.
In this conversion, the server first sends the SDRP codebooks to the clients.
Then, clients decrypt the SDRP codebooks and pack them into WOP and WRP ciphertexts.
After that, the server receives the WOP and WRP codebooks.
The number of WOP codebooks $N_{WOP}$ and WRP codebooks $N_{WRP}$ are as follows:
\begin{align}
\label{eq:WOP}
N_{WOP} &= \left\lceil \frac{n_{s} N_{C}}{n_{pd}} \right\rceil = \left\lceil \frac{n_{s} N_{C}}{\left\lceil \frac{N_{r}}{2 \left\lceil \frac{d}{n_{s}} \right\rceil} \right\rceil} \right\rceil \approx \left\lceil \frac{2 d^{\prime} N_{C}}{N_{r}} \right\rceil, \\
\label{eq:WRP}
N_{WRP} &= N_{C} N_{WOP}.
\end{align}
Therefore, WOP and WRP can effectively pack codebooks considering only the pseudo-dimension $d^{\prime}$ and codebook size $N_{C}$.
Especially if $d$ is divisible by $n_{s}$, the number of sub-dimensions $n_{s}$ has little effect on $N_{WOP}$.
Thus, WOP and WRP are much more effective data packing than SDOP and SDRP.
WOP and WRP are greatly beneficial in reducing computational complexity and network traffic after the codebook generation process.

\subsection{Database Encoding}
Database encoding assigns codebook indexes to data for each subspace.
Firstly, data are encrypted with WRP, which corresponds to WOP codebooks.
Database encoding simply computes distances between the WRP of data and the WOP of codebooks and assigns codebook indexes with the minimum distance to data.
The algorithm for database encoding for servers is presented in Algorithm~\ref{alg:ServerDatabaseEncoding}.
\begin{algorithm}[tb]
    \caption{Server algorithm of database encoding}
    \label{alg:ServerDatabaseEncoding}
    \textbf{Input:} WRP data ciphertexts $C_{x}$ and WOP codebook ciphertexts $C_{WOP}$ \\
    \textbf{Output:} Encoded data $\mathbf{x}_{E}$
    \begin{algorithmic}[1]
        \State {$C_{D} \leftarrow$ compute the distances between $C_{x}$ and $C_{WOP}$ with Algorithm~\ref{alg:RotSummation}}
        \State {Send distance ciphertexts $C_{D}$ to client}
        \State {$D \leftarrow$ receive decrypted distances from clients}
        \State {$\mathbf{x}^{(E)} \leftarrow$ assigns codebook index, which is minimum distance, for each subspace}
    \end{algorithmic}
\end{algorithm}
This algorithm computes distances of $N_{WOP}$ pairs only and communicates $2 N_{WOP}$ ciphertexts between the client and the server for each data.
Database encoding finally encodes the database $\mathcal{D}_{B} = \{ \mathbf{x}_{i} \}_{i=1}^{N_{B}} \in \mathbb{R}^{N_{B} \times d}$ into encoded database $\mathcal{D}_{E} = \{ \mathbf{x}^{(E)}_{i} \}_{i=1}^{N_{B}} \in \mathbb{Z}^{N_B \times n_{s}}$, where $N_{B}$ means database size, which $\mathcal{D}_{E}$ represents codebook indexes corresponding to the PQ codebooks for each subspace..
Note that the server stores $\mathcal{D}_{E}$ in plaintext, but the server can not restore the plaintext data because the codebooks are encrypted.
Additionally, this algorithm can be executed in parallel.
The server only stores an encoded database with a few bytes without the ciphertexts of the database.
All subsequent processing is performed using data represented by the PQ code.

\subsection{Database Indexing}
We use the inverted file (IVF) index \cite{PQ} as database indexing for PPPQ-ANN.
The IVF index needs a large number of centers $N_{I}$.
However, it is challenging to generate many centers using Algorithm~\ref{alg:ServerKMeans} and \ref{alg:ClientKMeans} due to the high computational cost and traffic.

PPPQ-ANN utilizes PQk-means \cite{PQkmeans} to generate a large number of centers $I_{E} = \{ \mathbf{c}_{i} \}_{i=1}^{N_{I}} \in \mathbb{Z}^{N_{I} \times n_{s}}$ from an encoded database $\mathcal{D}_{E}$.
Firstly, database indexing computes distances between codebooks for each subspace.
Then, database indexing performs PQk-means based on codebook distances.
After the generation of centers, each data point is assigned to $n_{nb}$ nearby centers based on asymmetric distance computation (ADC) \cite{PQ}.
The algorithm of database indexing is shown in Algorithm~\ref{alg:DatabaseIndexing}.
\begin{algorithm}[tb]
    \caption{Database indexing algorithm}
    \label{alg:DatabaseIndexing}
    \textbf{Input:} WOP codebook ciphertexts $C_{WOP}$, WRP codebook ciphertexts $C_{WRP}$, and encoded database $\mathcal{D}_{E}$ \\
    \textbf{Output:} IVF index $I_{IVF}$
    \begin{algorithmic}[1]
        \State {$C_{D} \leftarrow$ compute the distances between codebooks using $C_{WOP}$ and $C_{WRP}$ with Algorithm~\ref{alg:RotSummation}}
        \State {Send distance ciphertexts $C_{D}$ to client}
        \State {$D \leftarrow$ receive decrypted distances from clients}
        \State {$I_{E} \leftarrow$ perform PQk-means \cite{PQkmeans} with $D$ and $\mathcal{D}_{E}$}
        \State {$I_{IVF} \leftarrow$ for each $x^{(E)}_{i}$, find nearby $n_{nb}$ centers from $I_{E}$ and assign to those centers}
    \end{algorithmic}
\end{algorithm}
This algorithm only computes distances over the CKKS scheme between WOP codebook ciphertexts $C_{WOP}$ and WRP codebook ciphertexts $C_{WRP}$.
Therefore, the number of ciphertext computations is very small and can be computed only by the server during large-scale clustering.
Specifically, the number of ciphertext computations is $N_{C} N_{WOP}$, and the number of ciphertext traffic between servers and clients is also $2 N_{C} N_{WOP}$ ciphertexts.

The difference from the original IVF index is database encoding.
The original IVF index encodes the difference between a database and assigned centers.
This encoding contributes to a reduction of errors \cite{PQ}.
However, this encoding requires reencoding and communicating $2 N_{B} N_{WOP}$ ciphertexts between servers and clients.
Hence, PPPQ-ANN does not reencode and uses encoded data $\mathcal{D}_{E}$ for low complexity.

\subsection{Search Algorithm}
The search algorithm first computes distances between a query and codebooks.
Queries are encrypted with WRP, wchich corresponds to WOP codebooks.
Then, this algorithm searches data from the IVF index based on ADC \cite{PQ}.
The search algorithm is shown in Algorithm~\ref{alg:Search}.
\begin{algorithm}[tb]
    \caption{Search algorithm}
    \label{alg:Search}
    \textbf{Input:} WRP query ciphertexts $C_{Q}$, WRP codebook ciphertexts $C_{WRP}$, IVF index $I_{IVF}$, and search length $l$ \\
    \textbf{Output:} Search resulg $r$
    \begin{algorithmic}[1]
        \State {$C_{D} \leftarrow$ compute the distances between a query and codebooks using $C_{Q}$ and $C_{WRP}$ with Algorithm~\ref{alg:RotSummation}}
        \State {Send distance ciphertexts $C_{D}$ to client}
        \State {$D \leftarrow$ receive decrypted distances from clients}
        \State {$I \leftarrow$ search $l_{c}$ nearest indexes from $I_{IVF}$ with $D$ and ADC}
        \State {$r \leftarrow$ search $l$ nearest negighbors from $\mathcal{D}_{E}$ registered in $I$ with $D$ and ADC}
    \end{algorithmic}
\end{algorithm}
This algorithm also computes $N_{WOP}$ pairs and communicates $2 N_{WOP}$ ciphertexts in the same way as the database encoding algorithm for each query.

PPPQ-ANN can operate in a multi-client environment. Specifically, the database encoding and search processes must function in a multi-client setting.
To enable multi-client operation for these processes, PPPQ-ANN requires only the WOP codebook encrypted with a secret key for each client.
PPPQ-ANN makes it easy to add new clients by using Proxy Re-Encryption (PRE) \cite{PRE} or simply re-encryption with some noise for PQ codebooks.
Therefore, since this application involves multiple clients and a small number of servers, the processing speed of the servers is critical.

\subsection{Ciphertext Computational Complexity and Network Traffic}
Ciphertext computations have higher computational complexity than plaintext computations.
Additionally, ciphertext computations generate significant network traffic.
Ciphertext computational complexity and network traffic for each PPPQ-ANN step are shown in Table~\ref{tab:ComplexityTraffic}.
Database encoding and search are particularly important as they affect the ability to handle large-scale datasets and the speed of the service,respectively.
The computational complexity of database encoding and search is $O(N_{WOP})$.
$N_{WOP}$ is defined as equation~\eqref{eq:WOP}.
$N_{WOP}$ depends on data dimension $d$ and codebook size $N_{C}$, and is largely independent of the number of subdimensions $n_{s}$
In addition, PQ can represent $N_{p} = N_{C}^{n_{s}}$ patterns, and this number of patterns is more sensitive to $n_{s}$ than to $N_{C}$.
Therefore, PPPQ-ANN uses large $n_{s}$ and small $N_{C}$ to keep $N_{WOP}$ small.
Note that these parameter settings have a very small impact on codebook generation and database indexing.
Codebook generation cancels out the large $n_{s}$ and small $N_{C}$ because $N_{SDOP}=N_{C}$.
Database indexing reduces computational complexity due to the small $N_{C}$ and $N_{WOP}$.
\begin{table*}[tbp]
    \caption{Ciphertext computational complexity and network traffic. $B_{C}$ means a ciphertext bytes.}
    \label{tab:ComplexityTraffic}
    \centering{
        \begin{tabular}{c|c|c}
            \hline
            Step & Computational complexity & Network Traffic \\
            \hline
            Codebook generation & $O(n_{s} n_{k} N_{SDOP} N_{SDRP})$ & $n_{s} n_{k} N_{SDOP} N_{SDRP} B_{C}$ \\
            Database encoding (per data) & $O (N_{WOP})$ & $2 N_{WOP} B_{C}$ \\
            Database indexing & $O (N_{C} N_{WOP})$ & $2 N_{C} N_{WOP} B_{C}$ \\
            Search (per query) & $O (N_{WOP})$ & $2 N_{WOP} B_{C}$ \\
            \hline
        \end{tabular}
    }
\end{table*}

\section{Evaluation}
In this section, we validate the computational cost and network traffic for the PPPQ-ANN.
We provide two experiments.
The first experiment validates the performance in four benchmark datasets.
The second experiment validates the effect of the number of rotations $n_{rot}$ and sub dimensions $n_{s}$ on the performance.

\subsection{Datasets}
We evaluate PPPQ-ANN on three types of million-scale datasets: SIFT 1M \cite{PQ}, GIST 1M \cite{PQ}, and GloVe \cite{GloVe} datasets.
GloVe dataset has four different dimension datasets: 25 dimensions, 50 dimensions, 100 dimensions, and 200 dimensions.
Table~\ref{tab:Dataset} shows the details of the datasets.
Note that $\mathcal{D}$ and $\mathcal{D}_{B}$ are the same in the GloVe datasets.
\begin{table*}[tbp]
    \caption{
        Dataset details. 
        $\mathcal{D}$ denotes the training dataset used in codebook generation.
        $\mathcal{D}_{B}$ denotes the database dataset.
        $\mathcal{D}_{Q}$ denotes the query dataset.
        $d$ denotes the dataset dimension.
    }
    \label{tab:Dataset}
    \centering{
        \begin{tabular}{c|r|r|r|r|c}
            \hline
            Datasets & \multicolumn{1}{|c|}{$|\mathcal{D}|$} & \multicolumn{1}{|c|}{$|\mathcal{D}_{B}|$} & \multicolumn{1}{|c|}{$|\mathcal{D}_{Q}|$} & \multicolumn{1}{|c|}{$d$} & Metric\\
            \hline
            SIFT \cite{PQ} & 100,000 & 1,000,000 & 10,000 & 128 & Euclidean \\
            GIST \cite{PQ} & 500,000 & 1,000,000 & 1,000 & 960 & Euclidean \\
            GloVe-25d \cite{GloVe} & 1,183,514 & 1,183,514 & 10,000 & 25 & Inner Product \\
            GloVe-50d \cite{GloVe} & 1,183,514 & 1,183,514 & 10,000 & 50 & Inner Product \\
            GloVe-100d \cite{GloVe} & 1,183,514 & 1,183,514 & 10,000 & 100 & Inner Product \\
            GloVe-200d \cite{GloVe} & 1,183,514 & 1,183,514 & 10,000 & 200 & Inner Product \\
            \hline
        \end{tabular}
    }
\end{table*}

\subsection{Experimental Setup}
We implement PPPQ-ANN using C++ with the CKKS scheme implemented in the OpenFHE library \cite{OpenFHE, CKKS}.
This implementation also utilizes OpenMP.
All processes are implemented in a multi-process manner, except for search, which is implemented to process queries sequentially.
Table~\ref{tab:Parameters} shows the parameters of the CKKS scheme for each dataset type.
Note that PPPQ-ANN requires a multiplicative depth equal to 1 or more, and the optimal multiplicative depth is 1.
However, to prevent overflow, SIFT sets the multiplication depth to 2.
\begin{table*}[tbp]
    \caption{CKKS scheme parameters}
    \label{tab:Parameters}
    \centering{
        \begin{tabular}{c|c|c|c}
            \hline
            Parameters & SIFT & GIST & GloVe \\
            \hline
            Security level for secret key & 192bit & 192bit & 192bit\\
            Multiplicative depth & 2 & 1 & 1 \\
            Ring dimension $N_{r}$ & 16,384 & 16,384 & 16,384 \\
            Each ciphertext byte & 768KiB & 512KiB & 512KiB \\
            \hline
        \end{tabular}
    }
\end{table*}
This experiment uses two combinations of $n_{s}$ and $N_{C}$ for each dataset.
Table~\ref{tab:PQCode} shows the combination of $n_{s}$ and $N_{C}$, $N_{p}$, $N_{WOP}$, and the number of rotation functions $n_{rot}$ in Algorithm~\ref{alg:RotSummation}.
\begin{table*}[tbp]
    \caption{PQ division and codebook size}
    \label{tab:PQCode}
    \begin{center}
        \begin{tabular}{c|c|c|r|c|c|c}
            \hline
            Datasets & $n_{s}$ & $N_{C}$ & \multicolumn{1}{c|}{$\log_{2} (N_{p})$} & $d^{\prime}$ & $N_{WOP}$ & $n_{rot}$ \\
            \hline
            \multirow{2}{*}{SIFT} & 8 & 256 & 64 & 128 & 4 & 4 \\
            & 64 & 32 & 320 & 128 & 1 & 1 \\
            \hline
            \multirow{2}{*}{GIST} & 80 & 16 & 320 & 960 & 2 & 4 \\
            & 320 & 8 & 960 & 960 & 1 & 2 \\
            \hline
            \multirow{2}{*}{GloVe-25d} & 12 & 192 & 91 & 36 & 1 & 2 \\
            & 25 & 256 & 200 & 25 & 1 & 0 \\
            \hline
            \multirow{2}{*}{GloVe-50d} & 5 & 128 & 35 & 50 & 1 & 4 \\
            & 25 & 128 & 175 & 50 & 1 & 1 \\
            \hline
            \multirow{2}{*}{GloVe-100d} & 10 & 64 & 60 & 100 & 1 & 4 \\
            & 50 & 64 & 300 & 100 & 1 & 1 \\
            \hline
            \multirow{2}{*}{GloVe-200d} & 50 & 32 & 250 & 200 & 1 & 2 \\
            & 100 & 32 & 500 & 200 & 1 & 1 \\
            \hline
        \end{tabular}
    \end{center}
\end{table*}
Then, Table~\ref{tab:OtherParam} shows other parameters for PPPQ-ANN framework.
\begin{table*}[tbp]
    \caption{Other parameters for PPPQ-ANN framework}
    \label{tab:OtherParam}
    \begin{center}
        \begin{tabular}{c|c|c}
            \hline
            Step & Parameter & Value \\
            \hline
            \multirow{2}{*}{Codebook generation} & Number of samples used for secure k-means $N_{RS}$ & 1000 \\
            & Maximum iteration of k-means $n_{k}$ & 100 \\
            \hline
            \multirow{2}{*}{Database indexing} & IVF index centers $| I_{E} |$ & 1024 \\
            & Number of registrations for each data $n_{nb}$ & 3 \\
            \hline
            \multirow{2}{*}{Search} & Search length $l$ & 10 \\
            & Number of indexes to be searched $l_{c}$ & 3 \\
            \hline
        \end{tabular}
    \end{center}
\end{table*}
Evaluation metrics are Recall@10, server processing time $T_s$, client processing time $T_c$, server-side queries per second (QPS), total network traffic $\tau_{t}$, and network traffic per query $\tau_{q}$.
Additionally, we use the mean square error (MSE) for evaluating the quality of the codebook.
MSE is computed between the encoded database $\mathcal{D}_{E}$ and the original data in the database dataset $\mathcal{D}_{B}$.
Note that network traffic means the ciphertext traffic, excluding plaintext traffic.
All experiments are performed on a Google Cloud Compute Engine virtual instance.
The machine type for this instance is n2d-standard-32 with AMD-SEV-SNP enabled (32 AMD Milan vCPUs, 128GB of memory).

\subsection{The Results on Four Benchmark Datasets}
Tables~\ref{tab:CG}, \ref{tab:DE}, \ref{tab:IG}, and \ref{tab:Search} show the results of codebook generation, database encoding, index generation, and search processes, respectively.
For simplicity, we refer to the larger combination patterns (LCP) and the smaller combination patterns (SCP) according to the combination of PQ codebooks $N_{p}={N_{C}}^{n_{s}}$.
In each dataset of the result tables, the LCP corresponds to the lower results, and the SCP to the upper results.
\begin{table*}
    \caption{
        The results of the codebook generation process. 
        MSE denotes the computed between the encoded dataset $\mathcal{D}_{E}$ and the original data in the database dataset $\mathcal{D}_{B}$.
        $T_{s}$ and $T_{c}$ denote the server processing time and client processing time, respectively.
        $\tau_{t}$ denotes the amount of total traffic.
        QPS denotes the queries per second in database encoding.
        The better results in each dataset are shown in bold.
    }
    \label{tab:CG}
    \begin{center}
        \begin{tabular}{c|r|r||r|r|r|r}
            \hline
            \multirow{2}{*}{Datasets} & \multicolumn{2}{c||}{Parameter} & \multicolumn{4}{c}{Codebook Generation} \\
            \cline{2-7}
            & $n_{s}$ & $N_{C}$ & \multicolumn{1}{c|}{MSE} & \multicolumn{1}{c|}{$T_s$ [s]} & \multicolumn{1}{c|}{$T_c$ [s]} & \multicolumn{1}{c}{$\tau_{t}$[GiB]}  \\
            \hline \hline
            \multirow{2}{*}{SIFT} & 8 & 256 & 29.9 \texttimes 10\textsuperscript{3} & 1424.1 & 408.0 & 453.8 \\
                                  & 64 & 32 & \textbf{4.4 \texttimes 10\textsuperscript{3}} & \textbf{295.6} & \textbf{259.6} & \textbf{303.1} \\
            \hline
            \multirow{2}{*}{GIST} & 80 & 16 & 1.0 & \textbf{543.2} & \textbf{200.6} & \textbf{188.8} \\
                                  & 320 & 8 & \textbf{0.7} & 636.6 & 496.5 & 252.7 \\
            \hline
            \multirow{2}{*}{GloVe-25d} & 12 & 192 & \textbf{8.6} & 292.9 & \textbf{159.5} & \textbf{227.4} \\
                                       & 25 & 256 & 10.0 & \textbf{180.6} & 275.9 & 410.1 \\
            \hline
            \multirow{2}{*}{GloVe-50d} & 5 & 128 & 25.1 & \textbf{267.5} & \textbf{117.8} & \textbf{94.4} \\
                                       & 25 & 128 & \textbf{9.0} & 275.7 & 294.5 & 315.7 \\
            \hline
            \multirow{2}{*}{GloVe-100d} & 10 & 64 & 21.6 & 271.9 & \textbf{99.1} & \textbf{94.4} \\
                                        & 50 & 64 & \textbf{11.2} & \textbf{268.3} & 249.2 & 315.7 \\
            \hline
            \multirow{2}{*}{GloVe-200d} & 50 & 32 & 23.5 & \textbf{203.1} & \textbf{159.5} & \textbf{157.9} \\
                                        & 100 & 32 & \textbf{13.1} & 273.8 & 293.0 & 315.7 \\
            \hline
        \end{tabular}
    \end{center}
\end{table*}
\begin{table*}
    \caption{
        The results of the database encoding process. 
        $T_{s}$ and $T_{c}$ denote the server processing time and client processing time, respectively.
        $\tau_{q}$ denotes the amount of traffic per query.
        $\mathrm{QPS}_{s}$ denotes the server-side queries per second in database encoding.
        The better results in each dataset are shown in bold.
    }
    \label{tab:DE}
    \begin{center}
        \begin{tabular}{c|r|r||r|r|r|r}
            \hline
            \multirow{2}{*}{Datasets} & \multicolumn{2}{c||}{Parameter} & \multicolumn{4}{c}{Database Encoding} \\
            \cline{2-7}
            & $n_{s}$ & $N_{C}$ & \multicolumn{1}{c|}{$T_s$ [s]} & \multicolumn{1}{c|}{$T_c$ [s]} & \multicolumn{1}{c|}{QPS} & \multicolumn{1}{c}{$\tau_{q}$[MiB]} \\
            \hline \hline
            \multirow{2}{*}{SIFT} & 8 & 256 & 13897.4 & 2712.4 & 72.0 & 6.0 \\
                                  & 64 & 32 & \textbf{1411.2} & \textbf{659.9} & \textbf{708.6} & \textbf{1.5} \\
            \hline
            \multirow{2}{*}{GIST} & 80 & 16 & 4235.3 & 1059.5 & 236.1 & 2.0 \\
                                  & 320 & 8 & \textbf{1293.9} & \textbf{532.6} & \textbf{772.9} & \textbf{1.0} \\
            \hline
            \multirow{2}{*}{GloVe-25d} & 12 & 192 & 1498.9 & 628.8 & 789.6 & \textbf{1.0} \\
                                       & 25 & 256 & \textbf{522.4} & \textbf{625.5} & \textbf{2265.6} & \textbf{1.0} \\
            \hline
            \multirow{2}{*}{GloVe-50d} & 5 & 128 & 2507.0 & 844.0 & 472.1 & \textbf{1.0} \\
                                       & 25 & 128 & \textbf{1022.3} & \textbf{841.0} & \textbf{1157.7} & \textbf{1.0} \\
            \hline
            \multirow{2}{*}{GloVe-100d} & 10 & 64 & 2471.2 & \textbf{614.4} & 212.2 & \textbf{1.0} \\
                                        & 50 & 64 & \textbf{999.4} & 617.7 & \textbf{1184.2} & \textbf{1.0} \\
            \hline
            \multirow{2}{*}{GloVe-200d} & 50 & 32 & 1503.6 & 626.8 & 787.1 & \textbf{1.0} \\
                                        & 100 & 32 & \textbf{1019.1} & \textbf{630.8} & \textbf{1161.3} & \textbf{1.0} \\
            \hline
        \end{tabular}
    \end{center}
\end{table*}
\begin{table*}
    \caption{
        The results of the index generation process. 
        $T_{s}$ and $T_{c}$ denote the server processing time and client processing time, respectively.
        QPS denotes the queries per second in the search process.
        $\tau_{t}$ denotes the amount of total traffic.
        The better results in each dataset are shown in bold.
    }
    \label{tab:IG}
    \begin{center}
        \begin{tabular}{c|r|r||r|r|r}
            \hline
            \multirow{2}{*}{Datasets} & \multicolumn{2}{c||}{Parameter} & \multicolumn{3}{c}{Index Generation} \\
            \cline{2-6}
            & $n_{s}$ & $N_{C}$ & \multicolumn{1}{c|}{$T_s$[s]} & \multicolumn{1}{c|}{$T_c$[s]} & \multicolumn{1}{c}{$\tau_{t}$[MiB]} \\
            \hline \hline
            \multirow{2}{*}{SIFT} & 8 & 256 & \textbf{175.1} & 0.71 & 768.0 \\
                                  & 64 & 32 & 624.8 & 0.02 & \textbf{24.0} \\
            \hline
            \multirow{2}{*}{GIST} & 80 & 16 & \textbf{848.6} & 0.02 & 16.0 \\
                                  & 320 & 8 & 3757.8 & 0.03 & \textbf{4.0} \\
            \hline
            \multirow{2}{*}{GloVe-25d} & 12 & 192 & \textbf{345.6} & 0.10 & \textbf{96.0} \\
                                       & 25 & 256 & 307.3 & 0.14 & 128.0 \\
            \hline
            \multirow{2}{*}{GloVe-50d} & 5 & 128 & \textbf{156.6} & 0.10 & \textbf{64.0} \\
                                       & 25 & 128 & 317.0 & 0.10 & \textbf{64.0} \\
            \hline
            \multirow{2}{*}{GloVe-100d} & 10 & 64 & \textbf{350.0} & 0.03 & \textbf{32.0} \\
                                        & 50 & 64 & 771.7 & 0.04 & \textbf{32.0} \\
            \hline
            \multirow{2}{*}{GloVe-200d} & 50 & 32 & \textbf{767.2} & 0.02 & \textbf{16.0} \\
                                        & 100 & 32 & 1292.1 & 0.02 & \textbf{16.0} \\
            \hline
        \end{tabular}
    \end{center}
\end{table*}
\begin{table*}
    \caption{
        The results of the search process. 
        Recall@10 denotes the recall for a search length of 10.
        $T_{s}$ and $T_{c}$ denote the server processing time and client processing time, respectively.
        $\tau_{q}$ denotes the amount of traffic per query.
        QPS denotes the server-side queries per second in the search process.
        The better results in each dataset are shown in bold.
    }
    \label{tab:Search}
    \begin{center}
        \begin{tabular}{c|r|r||r|r|r|r|r}
            \hline
            \multirow{2}{*}{Datasets} & \multicolumn{2}{c||}{Parameter} & \multicolumn{5}{c}{Search} \\
            \cline{2-8}
            & $n_{s}$ & $N_{C}$ & \multicolumn{1}{c|}{Recall@10} & \multicolumn{1}{c|}{$T_s$[s]} & \multicolumn{1}{c|}{$T_c$[s]} & \multicolumn{1}{c|}{QPS} & \multicolumn{1}{c}{$\tau_{q}$[MiB]} \\
            \hline \hline
            \multirow{2}{*}{SIFT} & 8 & 256 & 99.6 & 591.5 & 110.4 & 16.9 & 6.0 \\
                                  & 64 & 32 & \textbf{100.0} & \textbf{249.5} & 120.6 & \textbf{40.1} & \textbf{1.5} \\
            \hline
            \multirow{2}{*}{GIST} & 80 & 16 & 87.1 & 45.4 & 13.4 & 22.0 & 2.0 \\
                                  & 320 & 8 & \textbf{94.1} & \textbf{30.289} & 12.5 & \textbf{33.0} & \textbf{1.0} \\
            \hline
            \multirow{2}{*}{GloVe-25d} & 12 & 192 & \textbf{90.9} & 249.7 & 80.4 & 40.0 & \textbf{1.0} \\
                                       & 25 & 256 & 66.7 & \textbf{136.0} & 97.8 & \textbf{73.5} & \textbf{1.0} \\
            \hline
            \multirow{2}{*}{GloVe-50d} & 5 & 128 & 44.0 & 384.7 & 81.0 & 26.0 & \textbf{1.0} \\
                                       & 25 & 128 & \textbf{97.7} & \textbf{186.9} & 97.4 & \textbf{53.5} & \textbf{1.0} \\
            \hline
            \multirow{2}{*}{GloVe-100d} & 10 & 64 & 71.3 & 373.2 & 80.4 & 26.8 & \textbf{1.0} \\
                                        & 50 & 64 & \textbf{97.9} & \textbf{183.7} & 87.0 & \textbf{54.4} & \textbf{1.0} \\
            \hline
            \multirow{2}{*}{GloVe-200d} & 50 & 32 & 92.0 & 251.0 & 86.9 & 39.8 & \textbf{1.0} \\
                                        & 100 & 32 & \textbf{97.1} & \textbf{179.6} & 108.8 & \textbf{55.7} & \textbf{1.0} \\
            \hline
        \end{tabular}
    \end{center}
\end{table*}

From Table~\ref{tab:CG}, the LCP generally had smaller MSEs, except for the GloVe-25d dataset, because combinations of PQ codebooks are with respect to the number of representable data.
The LCP of GloVe-25d could fail clustering with secure k-means because the sub-dimension $d_{s}$ is equal to 1.
Additionally, according to Table~\ref{tab:Search}, the Recall@10s for the ANN benchmarks were also higher with the smaller MSEs than with the larger MSEs.
Recall@10 for all datasets was higher than 90, and these results show PPPQ-ANN can retrieve nearest neighbors.
Therefore, the LCP, such as combinations of a large $n_{s}$ and a small $N_{C}$, can perform high retrieval performances. 

Another point of view, in the PPPQ-ANN processing time, the SCPs had shorter $T_{s}$ for the index generation process than the LCP.
On the other hand, the LCPs had shorter $T_{s}$ for the database encoding and search processes than the SCPs.
In the codebook generation process, the results on whether LCP or SCP reduced processing time varied across datasets.
These results are due to the PQ parameters, which are the number of subspaces $n_{s}$ and the codebook size per subspace $N_{C}$.
The number of subspaces $n_{s}$ is a factor of the sub-dimension $d_{s} = \lceil\frac{d}{n_{s}}\rceil$.
The computational complexity of Algorithm~\ref{alg:RotSummation} is Equation~\eqref{eq:DistanceComplexity}, and this depends on $d_{s}$.
Therefore, the larger the number of subspaces $n_{s}$, the shorter the distance computations per ciphertext are.
On the other hand, the codebook size per subspace $N_{C}$ is a factor of the number of WOP and WRP ciphertexts, as shown in Equations~\eqref{eq:WOP} and \eqref{eq:WRP}.
Therefore, a small $N_{C}$ contributes to reducing the number of computations with ciphertexts.
Reducing the number of computations with ciphertexts also reduces the ciphertext traffic. 
Thus, combinations of large $n_{s}$ and small $N_{C}$ can reduce computational time on ciphertexts.
The computational time on ciphertexts is critical for the PPPQ-ANN performance, such as the database encoding and the search processes.
Note that a large $n_{s}$ has an impact on the time of database generation, such as the codebook generation and the index generation processes.
However, database generation is not performed frequently, and the time spent on database generation relative to codebook and index generation is small.
Additionally, combinations of large $n_{s}$ and small $N_{C}$ can be LCP because ${N_{C}}^{n_{s}}$ is exponentially grows as $n_{s}$ increases.
Therefore, combinations of large $n_{s}$ and small $N_{C}$ are suitable for PPPQ-ANN, from both perspectives, processing time and ANN accuracy.
Note that $N_{C}$ should not be set to an excessively small value, and the maximum value for which $N_{WOP}$ equals 1 is suitable.

Finally, the server-side QPS for queries executed sequentially during the search process ranged from 30 to 55 in LCPs. 
Especially, PPPQ-ANN achieves more than 50 QPS when the multiplicative depth is 1, $d_{s}=1$, and $N_{WOP}=1$ without effect of exact dimension $d$.
While FHE applications typically suffer from server-side processing bottlenecks \cite{SANNS, GraSS}, PPPQ-ANN achieved very high processing speeds on the server side with optimized PQ parameters and data packing.
Furthermore, PPPQ-ANN can accelerate these results even more by processing queries in parallel.
In PPPQ-ANN, on the other hand, client-side processes are only for decryption. Hence, client-side processes are light.
For instance, in the search process for the GloVe-200d dataset, the average processing time per query for the client is 0.011 seconds.

\subsection{The Effect of Rotation Function and Sub-Dimensions on the Performance}
We conduct an additional experiment to confirm the impact of particularly important sub-dimensions $n_{s}$ on the PPPQ-ANN.
This experiment uses the Glove-200d dataset, with $N_{C}$ = 32 codebooks per subspace.
Additionally, the number of subspaces is $n_{s} = \{2, 5, 10, 20, 25, 50, 100\}$.
Table~\ref{tab:RotAblation} shows the results of the search process and the MSE of the PQ codebook on the GloVe-200d dataset.
The number of subspaces $n_{s}$ is a factor of the number of rotation functions $n_{rot}$ for distance computations and the combination of PQ codebooks $N_{p} = {N_{C}}^{n_{c}}$.
The smaller $n_{rot}$, the faster the distance calculation became, and the higher the QPS improved.
On the other hand, the larger ${N_{C}}^{n_{s}}$, the smaller the MSE became, and the Recall@10 increased.
From these results, maximizing the number of subspaces $n_{s}$ benefits both search accuracy and speed.
However, note that the sub-dimension $d_{s}$ must not be set to 1.
\begin{table*}
    \caption{
        The results of MSE for PQ codebooks and the search process on the GloVe-200d dataset.
        $n_{s}$ and $N_{C}$ denote the number of subspaces and codebooks per subspace, respectively.
        $n_{rot}$ denotes the number of rotation functions in the distance computations of Algorithm~\ref{alg:RotSummation}.
        MSE denotes the computed between the encoded dataset $\mathcal{D}_{E}$ and the original data in the database dataset $\mathcal{D}_{B}$.
        Recall@10 denotes the recall for a search length of 10.
        $T_{s}$ and $T_{c}$ denote the server processing time and client processing time, respectively.
        QPS denotes the server-side queries per second in the search process.
        $\tau_{q}$ denotes the amount of traffic per query.
        The better results in each dataset are shown in bold.
    }
    \label{tab:RotAblation}
    \begin{center}
        \begin{tabular}{r|r|r||r||r|r|r|r|r}
            \hline
            \multicolumn{3}{c||}{Parameter} & \multicolumn{1}{c||}{Codebook} & \multicolumn{5}{c}{Search} \\ \hline
            $n_{s}$ & $N_{C}$ & $n_{rot}$ & \multicolumn{1}{c||}{MSE}& \multicolumn{1}{c|}{Recall@10} & \multicolumn{1}{c|}{$T_s$[s]} & \multicolumn{1}{c|}{$T_c$[s]} & \multicolumn{1}{c|}{QPS} & \multicolumn{1}{c}{$\tau_{q}$[MiB]} \\
            \hline \hline
            2 & 32 & 8 & 35.4 & 0.0 & 5179.4 & 88.0 & 1.9 & 1.0 \\ \hline
            5 & 32 & 6 & 33.8 & 16.2 & 520.2 & 80.8 & 19.2 & 1.0 \\ \hline
            10 & 32 & 5 & 21.1 & 41.0 & 448.0 & 82.1 & 22.3 & 1.0 \\ \hline
            20 & 32 & 4 & 25.3 & 70.5 & 387.8 & 80.6 & 25.8 & 1.0 \\ \hline
            25 & 32 & 3 & 24.6 & 79.2 & 316.3 & 90.8 & 31.6 & 1.0 \\ \hline
            50 & 32 & 2 & 23.5 & 92.0 & 251.0 & 86.9 & 39.8 & 1.0 \\ \hline
            100 & 32 & 1 & \textbf{13.1} & \textbf{97.1} & \textbf{179.6} & 108.8 & \textbf{55.4} & 1.0 \\ \hline
        \end{tabular}
    \end{center}
\end{table*}


\section{Conclusion}
We have proposed a Privacy-Preserving Product-Quantization Approximate Nearest Neighbor (PPPQ-ANN) framework.
PPPQ-ANN combines product quantization (PQ) with the CKKS-based FHE and TEE.
PPPQ-ANN consistently performs all ANN operations, including codebook generation, index generation, and search, on FHE and TEE.
Consequently, the server can perform vector search without knowing the data, and PPPQ-ANN provides multi-layered security for data.
By employing specific PQ parameters and data packing to minimize FHE ciphertext computations, PPPQ-ANN achieves a query rate exceeding 50 QPS  across data dimensions and demonstrates high recall rates (Recall@10 > 0.9) in query sequential nearest-neighbor search.
Furthermore, the database generation process for PPPQ-ANN can be constructed in approximately 2 hours of computation for million-scale datasets.
With these capabilities, PPPQ-ANN has the potential to enable privacy-preserving, efficient, and scalable search services in a wide range of applications.

Finally, PPPQ-ANN has a risk of codebook leakage.
This risk arises when clients memorize the distance between queries and the codebook.
However, this risk can be mitigated by adding random noise to the computed distance results and rotating the ciphertexts. 
Then, a server subtracts the noise from the decrypted distances.
Additionally, this paper does not address attacks based on access patterns in the search process.
Future work will address these issues.

\section*{Acknowlegements}
This work was supported by J's Communication Co., Ltd. and EAGLYS Inc.

\bibliographystyle{ieeetr}
\bibliography{references}
\end{document}